\documentclass[preprint,amsmath,amssymb,aps]{revtex4-2}

\usepackage{graphicx}
\usepackage{dcolumn}
\usepackage{bm}
\usepackage{xcolor}
\usepackage{amsfonts}
\usepackage{graphics}
\usepackage{hyperref}
\usepackage{slashed}
\usepackage{babel}
\makeatletter

\begin{document}

\title{Higher-derivative $\mathcal{N}=1$ and $\mathcal{N}=2$ supersymmetric Maxwell-Chern-Simons theories at one loop in superspace}

\author{F. S. Gama}
\email{fgama@unifap.br}
\affiliation{Departamento de Ci\^{e}ncias Exatas e Tecnol\'{o}gicas, Universidade Federal do Amap\'{a}, 68903-419, Macap\'{a}, Amap\'{a}, Brazil}

\begin{abstract}
	We define a higher-derivative generalization of Maxwell-Chern-Simons theory in $\mathcal{N}=1$ and $\mathcal{N}=2$ superspaces. In particular, the chosen higher-derivative operator is a polynomial function of the d'Alembertian of arbitrary degree, and it is introduced exclusively in the gauge sector. The main goal is to explicitly compute the one-loop quantum corrections to the superfield effective potential for these theories. This is carried out by means of background field quantization in a higher-derivative $R_\xi$ gauge. The effective potential is obtained in closed form and expressed in terms of the roots of polynomial functions.
\end{abstract}

\maketitle
\newpage  

\section{Introduction}

The Maxwell-Chern-Simons (MCS) theory is an extension of three-dimensional Maxwell electrodynamics obtained through the inclusion of a Chern-Simons term. This gives rise to a new mechanism for generating mass for the electromagnetic field while preserving gauge symmetry. This is fundamentally different from the Higgs mechanism, since the mass emerges from a term of purely topological origin \cite{Siegel,Schonfeld,JT}. In addition to this property, the study of MCS theory is also motivated by its role as a useful theoretical laboratory for the investigation of dualities. Indeed, shortly after the formulation of the MCS theory, it was shown in \cite{TPN,DJ} that the self-dual and MCS models are dual to each other, that is, they provide distinct mathematical descriptions of the same physical content. Since then, several studies of this duality have been carried out involving different extensions of the MCS theory, such as noncommutative \cite{noncom}, Lorentz-violating \cite{Lorentzviol}, and supersymmetric ones \cite{SUSY-dual}.

At the same time, the study of higher-derivative (HD) field theories is a recurrent and relevant theme in physics. The main reason is that these types of theories perform better than standard ones with regard to classical and quantum divergences \cite{newtonsing,QG}. Indeed, this improved ultraviolet behavior also allows the use of higher covariant derivatives to regularize standard field theories \cite{regularization}. Additionally, HD theories arise naturally in different contexts, such as semiclassical gravity theories \cite{Shapiro}, string models \cite{string}, and the effective field theory approach \cite{EFT}. As for the MCS theory, there is a considerable body of literature on extensions of this theory involving higher derivatives \cite{HDMCS}. For example, issues related to stability and Ostrogradsky ghosts were analyzed in \cite{PDS} for an HD version of the MCS theory obtained by adding the standard Maxwell term to a HD Chern-Simons term. This theory was also studied in \cite{ANPRS}, where the authors investigated the properties of microcausality and perturbative unitarity up to one-loop order, while in \cite{MP} a Hamiltonian analysis of the theory was carried out in order to construct the gauge generator and identify the independent gauge symmetry.

In \cite{GNP} and \cite{GGNPS}, we formulated generic three-dimensional HD gauge theories in $\mathcal{N}=1$ and $\mathcal{N}=2$ superspaces, respectively, which, for appropriate choices of the HD operators, reduce to supersymmetric generalizations of the HD MCS theory. For these generic theories, we explicitly calculated the one-loop superfield effective potential (EP), due to its importance as a theoretical instrument for studying spontaneous symmetry breaking induced by radiative corrections \cite{breaking}, the restoration of symmetry at high temperatures \cite{restoration}, and the decay of the false vacuum \cite{falsevacuum}. However, a major limitation of Ref. \cite{GNP} is that we did not find a final expression for the EP that applies to an HD MCS theory, while Ref. \cite{GGNPS} also has a limitation because, although it applies to an HD MCS theory, the final EP was obtained in an integral representation rather than in closed form in terms of elementary or standard special functions. Due to these limitations, our goals in the present work are (i) to determine the one-loop superfield EP for an HD MCS theory in $\mathcal{N}=1$ superspace, thereby filling this gap in the literature, and (ii) to obtain, for the first time, the one-loop superfield EP in closed form and expressed in terms of the roots of polynomial functions for an HD MCS theory in $\mathcal{N}=2$ superspace. To achieve these goals, we perform the calculations in an HD $R_\xi$ gauge and obtain the EP by evaluating the functional traces directly. This methodology is completely different from the one used in \cite{GNP,GGNPS}, where the calculations were performed in an HD Fermi-Feynman gauge and the EP was obtained by evaluating one-loop supergraphs.

The paper is organized as follows. In Sec. \ref{secII}, we define an HD MCS theory in $\mathcal{N}=1$ superspace and compute its one-loop superfield EP. In Sec. \ref{secIII}, we define an HD MCS theory in $\mathcal{N}=2$ superspace and compute its one-loop superfield EP. Finally, in Sec. \ref{secIV}, we present a brief summary of the main results obtained and suggest a possible continuation of this study.

\section{HD MCS theory in $\mathcal{N}=1$ superspace}\label{secII}

In the $\mathcal{N}=1$ superfield formalism, the MCS theory is a $U(1)$ gauge theory described by an unconstrained spinor superfield $A_\alpha$ \cite{Siegel,GGRS}. We assume that this superfield interacts with massless charged matter, described by a complex superfield $\Phi$, through a minimal coupling. In this work, we study the following HD generalization of this theory:
\begin{equation}
	\label{model_1}
	S_{\mathcal{N}=1}=\frac{1}{2g^2}\int d^5z\left[W^\alpha f(\Box)W_\alpha+mA^\alpha f(\Box)W_\alpha\right]+\int d^5z\bar\Phi\nabla^2\Phi,
\end{equation}
where $W_\alpha=\frac{1}{2}D^\beta D_\alpha A_\beta$ is the gauge-invariant field strength. The scalar operator $f(\Box)$ is assumed to be a polynomial in the d'Alembertian of arbitrary degree. Also, in order to recover the usual MCS theory coupled to matter, we assume that $f(\Box)$ tends to $1$ in a suitable limit. It is worth mentioning that, in principle, we could also introduce HD operators in the matter sector and add self-interaction terms for the matter superfields to (\ref{model_1}). However, this would make the calculations more laborious and would increase the difficulty of evaluating the functional trace that will be computed later. Thus, for the sake of simplicity, we restrict our study to the model defined in Eq. (\ref{model_1}).

In the background field approach \cite{HSKS,GSR}, we split the original superfields $A_\alpha$ and $\Phi$ into background and quantum superfields through the substitution
\begin{equation}
	\label{split}
	A_\alpha\rightarrow A_\alpha+a_\alpha; \ \ \ \Phi\rightarrow\Phi+\phi; \ \ \ \bar\Phi\rightarrow\bar\Phi+\bar\phi
\end{equation}
into the action (\ref{model_1}). Note that the splitting (\ref{split}) is linear because (\ref{model_1}) is invariant under linear infinitesimal gauge transformations. The quantum superfields $a_\alpha$ and $\phi$ are unconstrained, whereas the classical background superfields $A_\alpha$ and $\Phi$ are required to satisfy
\begin{equation}
	\label{constr}
	A_\alpha=0; \ \ \ D_\alpha\Phi=0; \ \ \ D_{\alpha}\bar{\Phi}=0; \ \ \ \partial_{\alpha\beta}\Phi=0; \ \ \ \partial_{\alpha\beta}\bar{\Phi}=0.
\end{equation}
These constraints follow from the definition of the superfield EP, which depends only on the background matter superfield \cite{FGLNPSS}.

Therefore, upon substituting Eqs. (\ref{split}) into (\ref{model_1}), we obtain
\begin{equation}
	\label{splited_1}
	S_{\text{split}1}=\frac{1}{2g^2}\int d^5z\left[w^\alpha f(\Box)w_\alpha+ma^\alpha f(\Box)w_\alpha\right]+\int d^5z\left(\bar\Phi+\bar\phi\right)\nabla^2\left(\Phi+\phi\right),
\end{equation}
where $w_\alpha=\frac{1}{2}D^\beta D_\alpha a_\beta$.

The invariance of (\ref{model_1}) under $U(1)$ transformations implies that (\ref{splited_1}) is invariant under the following two types of transformations:
\begin{align}
	\label{background}
	\text{Background:} \ \ &\Phi^\prime=e^{i\omega}\Phi; \ \ \bar{\Phi}^\prime=e^{-i\omega}\bar{\Phi}; \ \ \phi^\prime=e^{i\omega}\phi; \ \ \bar{\phi}^\prime=e^{-i\omega}\bar{\phi}; \ \ a_\alpha^\prime=a_\alpha;\\
	\label{quantum}
	\text{Quantum:} \ \ &\Phi^\prime=\Phi; \ \ \bar{\Phi}^\prime=\bar{\Phi}; \ \ \phi^\prime=e^{ iK}\left(\Phi+\phi\right)-\Phi; \ \ \bar{\phi}^\prime=e^{-iK}\left(\bar{\Phi}+\bar{\phi}\right)-\bar{\Phi};\nonumber\\
	&a_\alpha^\prime=a_\alpha+D_\alpha K.
\end{align}
where $\omega$ and $K$ are the global and local transformation parameters, respectively. The background transformations are global because the background gauge superfield has been set to zero.

For the computation of the one-loop correction to the EP, it is sufficient to expand (\ref{splited_1}) around the background superfields and retain only the terms that are quadratic in the quantum superfields. By following this prescription, we obtain
\begin{equation}
	\begin{split}
		\label{quadratic}
		S^{(2)}_1=&\int d^5z\bigg\{\frac{1}{4g^2}a^\alpha\left[\left(\Box f(\Box)-2g^2\left|\Phi\right|^2\right){\delta_\alpha}^\beta+f(\Box)i{\partial_\alpha}^\beta D^2+mf(\Box)D^\beta D_\alpha\right]a_\beta\\
		&+\bar{\phi}D^2\phi-\frac{i}{2}\bar \Phi a^\alpha D_\alpha\phi+\frac{i}{2}\Phi a^\alpha D_\alpha\bar\phi\bigg\}.
	\end{split}
\end{equation}
For our purposes, it is convenient to choose the following parametrization for the complex matter superfields:
\begin{equation}
	\Phi=\frac{1}{\sqrt{2}}\left(\Phi_1+i\Phi_2\right) \ \ ; \ \ \phi=\frac{1}{\sqrt{2}}\left(\phi_1+i\phi_2\right),
\end{equation}
where $\Phi_1,\Phi_2$ and $\phi_1,\phi_2$ are real superfields. Thus, Eq. (\ref{quadratic}) must be rewritten as
\begin{equation}
	\begin{split}
		\label{quad_real}
		S^{(2)}_1=&\int d^5z\bigg\{\frac{1}{4g^2}a^\alpha\left[\left(\Box f(\Box)-2M^2_A\right){\delta_\alpha}^\beta+f(\Box)i{\partial_\alpha}^\beta D^2+mf(\Box)D^\beta D_\alpha\right]a_\beta\\
		&+\frac{1}{2}\phi_n D^2\phi_n+\frac{1}{2}\varepsilon_{nm}\Phi_n a^\alpha D_\alpha\phi_m\bigg\},
	\end{split}
\end{equation}
where $n,m\in\{1,2\}$ and $\varepsilon_{nm}$ is the Levi-Civita symbol. 
Moreover, the mass parameter $M_A^2$ is defined by
\begin{equation}
	\label{mass_gauge}
	M_A^2:=\frac{g^2}{2}\Phi_n\Phi_n=g^2\left|\Phi\right|^2.
\end{equation}
It is also convenient to eliminate from (\ref{quad_real}) the mixed term proportional to $a^\alpha D_\alpha\phi_m$. One way to achieve this would be to perform a nonlocal change of variables in the path integral associated with (\ref{quad_real}). However, in the present work, we decouple the quantum superfields $a^\alpha$ and $\phi_m$ by fixing the quantum gauge symmetry (\ref{quantum}). Indeed, the unwanted mixed terms can be removed by adding to (\ref{quad_real}) the following HD generalization of the supersymmetric $R_\xi$ gauge:
\begin{equation}
	\label{gft_1}
	S_{GF1}=-\frac{1}{g^2}\int d^5zFf(\Box)D^2F,
\end{equation}
where the gauge-fixing function is defined by
\begin{equation}
	\label{gff_1}
	F:=\frac{1}{2}\left(D^\alpha a_\alpha-g^2\Phi_n\varepsilon_{nm}\frac{D^2}{\Box f(\Box)}\phi_m\right).
\end{equation}
This gauge-fixing function is invariant under global background transformations and, therefore, so is $S_{GF1}$. Now, by substituting (\ref{gff_1}) into (\ref{gft_1}), expanding the result, and integrating by parts, we arrive at
\begin{equation}
	\label{gfft_1}
	\begin{split}
		S_{GF1}=&\int d^5z\bigg[\frac{1}{4g^2}a^\alpha\Box f(\Box)a_\alpha-\frac{1}{4g^2}a^\alpha f(\Box)i{\partial_\alpha}^\beta D^2a_\beta-\frac{1}{2}\Phi_n\varepsilon_{nm}a^\alpha D_\alpha\phi_m\\
		&-\frac{1}{2}M_{nm}^2\phi_n\frac{D^2}{\Box f(\Box)}\phi_m\bigg],
	\end{split}
\end{equation}
where $M_{nm}^2$ denotes the entries of the mass matrix for the scalar multiplet, defined as
\begin{equation}
	M^2_{nm}:=\frac{g^2}{2}\left(\Phi_\ell\Phi_\ell\delta_{nm}-\Phi_n\Phi_m\right). 
\end{equation}
Now, since the gauge-fixing function (\ref{gff_1}) changes under infinitesimal quantum gauge transformations according to
\begin{equation}
	\delta_K F=D^2K-M^2_A\frac{D^2}{\Box f(\Box)}K-\frac{g^2}{2}\Phi_n\frac{D^2}{\Box f(\Box)}\left(K\phi_n\right),
\end{equation}
it follows that the Faddeev-Popov ghosts interact with the background superfields at the one-loop level, so that their contribution is, in principle, non-trivial. Therefore, we must also add to (\ref{quad_real}) the ghost action
\begin{equation}
	\label{ghost}
	S_{FP1}=\left.\int d^5z\tilde c\delta_K F\right|_{K\to c}=\int d^5z\tilde c\left(1-\frac{M^2_A}{\Box f(\Box)}\right)D^2c,
\end{equation}
where we have disregarded the cubic term because it does not contribute to the one-loop EP.

Finally, the total action quadratic in the quantum superfields, which includes the gauge-fixing and ghost terms, is obtained by adding (\ref{gfft_1}) and (\ref{ghost}) to (\ref{quadratic}). Therefore,
\begin{equation}
	\label{combined}
	\begin{split}
		S^{(2)}_1&+S_{GF1}+S_{FP1}=\frac{1}{2}\int d^5zd^5z^\prime \Bigg[a^\alpha(z){\left(\mathcal{H}_a\right)_\alpha}^\beta\delta^5(z,z^\prime) a_\beta(z^\prime)\\
		&+\phi_n(z)\left(\mathcal{H}_\phi\right)_{nm}\delta^5(z,z^\prime)\phi_m(z^\prime)+\left(\begin{array}{cc}
			c(z) \ \ \ &
			\tilde c(z)\end{array}\right)\mathcal{H}_{FP1}\delta^5(z,z^\prime)\left(\begin{array}{c}
			c(z^\prime)\\
			\tilde c(z^\prime)\end{array}\right)\Bigg],
	\end{split}
\end{equation}
where the Hessians are given by
\begin{align}
	\label{H_a}
	{\left(\mathcal{H}_a\right)_\alpha}^\beta&=\frac{1}{g^2}\left[\left(\Box f(\Box)-M^2_A\right){\delta_\alpha}^\beta+\frac{m}{2}f(\Box)D^\beta D_\alpha\right];\\
	\label{H_phi}
	\left(\mathcal{H}_\phi\right)_{nm}&=\left[\delta_{nm}-\frac{M^2_{nm}}{\Box f(\Box)}\right]D^2;\\
	\label{H_FP1}
	\mathcal{H}_{FP1}&=\left(1-\frac{M^2_A}{\Box f(\Box)}\right)\left(-i\sigma_2\right)D^2.
\end{align}
Note that, as a consequence of the gauge choice (\ref{gft_1}), the Hessian corresponding to the matter superfield, given in (\ref{H_phi}), is also modified by the HD operator $f(\Box)$, even though we have not introduced it into the matter sector of (\ref{model_1}).

The Hessians given above are essential for determining the one-loop correction to the effective action, since the calculation ultimately reduces to computing the trace of the logarithm of the Hessians above. Indeed, by integrating out the quantum superfields, it can be shown that the one-loop Euclidean effective action is given by
\begin{equation}
	\label{1loopEA_1}
	\Gamma^{(1)}_{\mathcal{N}=1}=\frac{1}{2}\textrm{Tr}\ln{\left(\mathcal{H}_a\right)_\alpha}^\beta-\frac{1}{2}\textrm{Tr}\ln\left(\mathcal{H}_\phi\right)_{nm}+\frac{1}{2}\textrm{Tr}\ln\mathcal{H}_{FP1}.
\end{equation}
The second and third traces can be disregarded for the following reason. Let
\begin{equation}
	\label{Cs}
	\left(\mathcal{C}_\phi\right)_{nm}=\delta_{nm}f(\Box)D^2 \ \ ; \ \ \mathcal{C}_{FP1}=i\sigma_2f(\Box)D^2
\end{equation}
be operators. Since they are independent of the background superfields, adding $-\frac{1}{2}\textrm{Tr}\ln\left(\mathcal{C}_\phi\right)_{nm}$ or $-\frac{1}{2}\textrm{Tr}\ln\mathcal{C}_{FP1}$ to (\ref{1loopEA_1}) does not modify the result, because these quantities are integrals of a constant function, which vanish due to the properties of the Berezin integral. Thus, we may add them to (\ref{1loopEA_1}) and use Eqs. (\ref{H_phi}), (\ref{H_FP1}), and (\ref{Cs}) to obtain the following traces:
\begin{align}
	\label{trivial_1}
	-\frac{1}{2}\textrm{Tr}\ln\left(\mathcal{C}_\phi\mathcal{H}_\phi\right)_{nm}=&-\frac{1}{2}\textrm{Tr}\ln\left(\delta_{nm}\Box f(\Box)-M^2_{nm}\right);\\
	\label{trivial_2}
	\frac{1}{2}\textrm{Tr}\ln\left(\mathcal{C}_{FP1}\mathcal{H}_{FP1}\right)=&\frac{1}{2}\textrm{Tr}\ln\left(\left(\Box f(\Box)-M^2_A\right)I_{2\times2}\right).
\end{align}
Even though these are traces of operators that depend on the background superfields, we can drop them because the operators contain no spinor covariant derivatives, which implies that the traces in Eqs. (\ref{trivial_1}) and (\ref{trivial_2}) vanish.

Therefore, only the first trace in (\ref{1loopEA_1}) is non-trivial, and by using (\ref{H_a}) we can write it as
\begin{equation}
	\label{1loopEA_1_2}
	\Gamma^{(1)}_{\mathcal{N}=1}=\frac{1}{2}\textrm{Tr}\ln\frac{1}{g^2}\left[\left(\Box f(\Box)-M^2_A\right){\delta_\alpha}^\beta+\frac{m}{2}f(\Box)D^\beta D_\alpha\right].
\end{equation}
Note that, if $m=0$, this one-loop correction vanishes. This means that the Chern-Simons term in (\ref{model_1}) is essential for ensuring that the one-loop EP is non-trivial.

We can rewrite Eq. (\ref{1loopEA_1_2}) as follows:
\begin{align}
	\Gamma^{(1)}_{\mathcal{N}=1}&=\frac{1}{2}\textrm{Tr}\ln\left[\frac{1}{g^2}\left(\Box f(\Box)-M^2_A\right){\delta_\alpha}^\gamma\right]+\frac{1}{2}\textrm{Tr}\ln\left[{\delta_\gamma}^\beta+\frac{m}{2}\frac{f(\Box)}{\Box f(\Box)-M^2_A}D^\beta D_\gamma\right]\nonumber\\
	&=\frac{1}{2}\textrm{Tr}\ln\left[{\delta_\gamma}^\beta+\frac{m}{2}\frac{f(\Box)}{\Box f(\Box)-M^2_A}D^\beta D_\gamma\right]\nonumber\\
	&=-\frac{1}{2}\int d^5z\int \frac{d^3p}{(2\pi)^3}\frac{1}{p}\arctan\left(\frac{mpf(-p^2)}{p^2f(-p^2)+M^2_A}\right),
\end{align}
where the last equality follows from Eq. (16) of \cite{GNP2}. Since this result was obtained under the assumption that the background matter superfields are constant, it follows immediately that the one-loop EP is given by
\begin{equation}
	\label{pot_integral}
	K^{(1)}_{\mathcal{N}=1}=-\frac{1}{2}\int \frac{d^3p}{(2\pi)^3}\frac{1}{p}\arctan\left(\frac{mpf(-p^2)}{p^2f(-p^2)+M^2_A}\right).
\end{equation}
The last step consists of evaluating the Euclidean Feynman integral. We can handle the integral in (\ref{pot_integral}) by defining the vector field
\begin{equation}
	V_\mu(p):=\frac{p_\mu}{p}\arctan\left(\frac{mpf(-p^2)}{p^2f(-p^2)+M^2_A}\right)
\end{equation}
and using the fact that, in dimensional regularization, surface terms can be disregarded:
\begin{equation}
	\mu^{2\varepsilon}\int\frac{d^dp}{(2\pi)^d}\frac{\partial V_\mu(p)}{\partial p_\mu}=0.
\end{equation}
After some algebraic manipulations, this identity leads to
\begin{equation}
	\begin{split}
		\mu^{2\varepsilon}\int\frac{d^dp}{(2\pi)^d}\frac{1}{p}\arctan&\left(\frac{mpf(-p^2)}{p^2f(-p^2)+M^2_A}\right)=\\
		&-\frac{\mu^{2\varepsilon}}{d-1}\int\frac{d^dp}{(2\pi)^d}\frac{d}{dp}\arctan\left(\frac{mpf(-p^2)}{p^2f(-p^2)+M^2_A}\right).
	\end{split}
\end{equation}
Substituting this expression into (\ref{pot_integral}) and taking the derivative, we obtain
\begin{equation}
	\label{pot_integral_x}
	K^{(1)}_{\mathcal{N}=1}=-\frac{m\mu^{2\varepsilon}}{2(d-1)}\int\frac{d^dp}{(2\pi)^d}\frac{N(s)}{D(s)},
\end{equation}
where $s:=p^2$ and
\begin{align}
	\label{num}
	N(p^2)&:=p^2f^2(-p^2)-M^2_A\left[f(-p^2)+p\frac{d}{dp}f(-p^2)\right];\\
	\label{den}
	D(p^2)&:=\left[p^2f(-p^2)+M^2_A\right]^2+m^2p^2f^2(-p^2).
\end{align}
Suppose that the coefficients of the polynomial $f(\Box)$ are chosen in such a way that $D(s)$ has distinct roots $s_1,s_2,\ldots,s_n$. Since $N(s)$ has degree lower than that of $D(s)$, we can represent the rational function in (\ref{pot_integral_x}) by means of a partial fraction decomposition \cite{Ahlfors}:
\begin{equation}
	\frac{N(s)}{D(s)}=\sum_{j=1}^n\frac{N(s_j)}{D^\prime(s_j)}\frac{1}{s-s_j}.
\end{equation}
Plugging this expression into (\ref{pot_integral_x}), we obtain a sum of Feynman integrals that can be evaluated exactly. Therefore, the one-loop superfield effective potential for the $\mathcal{N}=1$ higher-derivative Maxwell-Chern-Simons theory (\ref{model_1}) is given by
\begin{equation}
	\label{final_1}
	K^{(1)}_{\mathcal{N}=1}=\frac{m}{16\pi}\sum_{j=1}^n\frac{\sqrt{-s_j}\,N(s_j)}{D^\prime(s_j)}.
\end{equation}
This result makes explicit how the functional structure of the one-loop correction is governed by the degrees of freedom of the HD theory. In fact, the higher the degree of the polynomial $f(\Box)$ in (\ref{model_1}), the larger the number of extra propagating degrees of freedom, including, in general, Ostrogradsky ghosts. At the same time, as the degree of $f(\Box)$ increases, new roots $s_j$ appear, and the one-loop correction acquires additional contributions of the form $\sqrt{-s_j}\,N(s_j)/D^\prime(s_j)$.

As an application of our general result, let us consider, for the sake of simplicity, a theory without higher derivatives, that is, $f(\Box)=1$. This assumption, in conjunction with Eqs. (\ref{num}) and (\ref{den}), implies
\begin{equation}
	\label{poly}
	N(s)=s-g^2|\Phi|^2 \ \ ; \ \ D(s)=\left(s+g^2|\Phi|^2\right)^2+m^2s=(s-s_+)(s-s_-),
\end{equation}
where we have used (\ref{mass_gauge}). The roots of $D(s)$ are given by
\begin{equation}
	\label{roots}
	s_\pm=-\frac{1}{2}\left(m^2+2g^2|\Phi|^2\pm m^2\sqrt{1+\frac{4g^2|\Phi|^2 }{m^2}}\right),
\end{equation}
and these roots are negative real numbers. Finally, the one-loop potential follows from (\ref{final_1}) and (\ref{poly}):
\begin{equation}
	\left.K_{\mathcal{N}=1}^{(1)}\right|_{f(\Box)=1}=\frac{m}{16\pi}\left[\frac{\sqrt{-s_+}\left(s_+-g^2|\Phi|^2\right)}{s_+-s_-}+\frac{\sqrt{-s_-}\left(s_--g^2|\Phi|^2\right)}{s_--s_+}\right],
\end{equation}
where $s_+$ and $s_-$ are given in (\ref{roots}).

Of course, we could also consider a HD operator, for example, $f(\Box)=1+\frac{\Box}{\Lambda^2}$. However, in this case, the polynomial $D(s)$ has degree four, and its roots are very cumbersome to write out explicitly in full. That is why we have left this example out of the present discussion.

\section{HD MCS theory in $\mathcal{N}=2$ superspace}\label{secIII}

In contrast to the MCS theory in $\mathcal{N}=1$ superspace, the corresponding theory in $\mathcal{N}=2$ superspace is described by a real scalar superfield $V$. We assume a minimal coupling between this gauge superfield and a pair of massless chiral matter superfields $\Phi_+$ and $\Phi_-$ carrying non-zero $U(1)$ charges \cite{N2SQED}. For a HD generalization of this theory, we introduce HD terms only in the gauge sector, as follows
\begin{equation}
	\label{model_2}
	S_{\mathcal{N}=2}=\int d^7z\left(-\frac{1}{8g^2}Gf(\Box)G+\frac{m}{8g^2}Vf(\Box)G+\bar{\Phi}_+e^V\Phi_++\bar{\Phi}_- e^{-V}\Phi_-\right),
\end{equation}
where the gauge-invariant field strength is now the scalar superfield $G=\bar{D}^\alpha D_\alpha V$. The dimensionless scalar operator $f(\Box)$ is the same as the one defined in (\ref{model_1}). Again, for the sake of simplicity, we have neither included HD operators in the matter sector nor added to (\ref{model_2}) self-interaction terms involving the chiral superfields.

Using the background field approach, let us linearly split the superfields in (\ref{model_2}) into background $\{V,\Phi_+,\Phi_-\}$ and quantum $\{v,\phi_+,\phi_-\}$ parts as follows:
\begin{equation}
	\label{split_2}
	V\rightarrow V+v; \ \ \ \Phi_\pm\rightarrow\Phi_\pm+\phi_\pm; \ \ \ \bar{\Phi}_\pm\rightarrow\bar{\Phi}_\pm+\bar{\phi}_\pm.
\end{equation}
Since, by definition, the $\mathcal{N}=2$ K\"{a}hlerian EP depends only on chiral and antichiral superfields, but not on their derivatives \cite{BKY}, we assume that the background superfields are subject to the following constraints:
\begin{equation}
	\label{constr_2}
	V=0; \ \ \ D_\alpha\Phi_\pm=0; \ \ \ \bar{D}_{\beta}\bar{\Phi}_\pm=0; \ \ \ \partial_{\alpha\beta}\Phi_\pm=0; \ \ \ \partial_{\alpha\beta}\bar{\Phi}_\pm=0.
\end{equation}
Therefore, upon inserting Eqs. (\ref{split_2}) into (\ref{model_2}), we obtain
\begin{equation}
	\begin{split}
		\label{splited_2}
		S_{split2}&=\int d^7z\bigg[-\frac{1}{8g^2}gf(\Box)g+\frac{m}{8g^2}vf(\Box)g+\left(\bar{\Phi}_++\bar{\phi}_+\right)e^v\left(\Phi_++\phi_+\right)\\
		&+\left(\bar{\Phi}_-+\bar{\phi}_-\right)e^{-v}\left(\Phi_-+\phi_-\right)\bigg],
	\end{split}
\end{equation}
where $g=\bar{D}^\alpha D_\alpha v$. It is straightforward to verify that (\ref{splited_2}) is invariant under the following two kinds of transformations:
\begin{align}
	\label{background_2}
	\text{Background:} \ \ &\Phi_{\pm}^\prime=e^{\pm iK}\Phi_{\pm}; \ \ \bar{\Phi}_{\pm}^\prime=e^{\mp iK}\bar{\Phi}_{\pm}; \ \ \phi_{\pm}^\prime=e^{\pm iK}\phi_{\pm}; \ \ \bar{\phi}_{\pm}^\prime=e^{\mp iK}\bar{\phi}_{\pm}; \ \ v^\prime=v;\\
	\label{quantum_2}
	\text{Quantum:} \ \ &\Phi_{\pm}^\prime=\Phi_{\pm}; \ \ \bar{\Phi}_{\pm}^\prime=\bar{\Phi}_{\pm}; \ \ \phi_{\pm}^\prime=e^{\pm i\Lambda}\left(\Phi_{\pm}+\phi_{\pm}\right)-\Phi_{\pm};\nonumber\\
	&\bar{\phi}_{\pm}^\prime=e^{\mp i\bar{\Lambda}}\left(\bar{\Phi}_{\pm}+\bar{\phi}_{\pm}\right)-\bar{\Phi}_{\pm}; \ \ v^\prime=v+i\left(\bar{\Lambda}-\Lambda\right),
\end{align}
where $\Lambda$ and $\bar{\Lambda}$ are local (anti)chiral parameters, and $K$ is a global real parameter.

Since we are interested in the one-loop EP, only terms up to second order in the quantum superfields are relevant. Therefore, we have
\begin{equation}
	\begin{split}
		\label{quadratic_2}
		S^{(2)}_2=\int d^7z\bigg\{&\frac{1}{8g^2}v\left[f(\Box)\left(-\Box+\{D^2,\bar{D}^2\}\right)+mf(\Box)\bar D^\alpha D_\alpha+M^2_v\right]v\\
		&+\bm{\bar{\phi}}\bm{\phi}+v\bm{\bar{\Phi}}\sigma_3\bm{\phi}+v\bm{\bar{\phi}}\sigma_3\bm{\Phi}\bigg\},
	\end{split}
\end{equation}
where we have introduced the following matrix notation:
\begin{equation}
	\label{matrix_not}
	\bm{\bar{\Phi}}=\left(\begin{array}{cc}
		\bar{\Phi}_+ & \bar{\Phi}_- 
	\end{array}\right); \ \ \ \bm{\Phi}=\left(\begin{array}{c}
		\Phi_+ \\
		\Phi_-
	\end{array}\right); \ \ \ \bm{\bar{\phi}}=\left(\begin{array}{cc}
		\bar{\phi}_+ & \bar{\phi}_- 
	\end{array}\right); \ \ \ \bm{\phi}=\left(\begin{array}{c}
		\phi_+ \\
		\phi_-
	\end{array}\right). 
\end{equation}
Moreover, $\sigma_3$ denotes a Pauli matrix, and the mass parameter $M_v^2$ is defined by
\begin{equation}
	\label{mass_gauge_2}
	M_v^2:=4g^2\bm{\bar{\Phi}}\bm{\Phi}.
\end{equation}
Similarly to what we have done in the previous section, we now remove the mixed terms $v\bm{\bar{\Phi}}\sigma_3\bm{\phi}$ and $v\bm{\bar{\phi}}\sigma_3\bm{\Phi}$ from (\ref{quadratic_2}) by adding to it the following HD $R_\xi$ gauge-fixing term:
\begin{equation}
	\label{gft_2}
	S_{GF2}=-\frac{1}{4g^2}\int d^7z\bar{F}f(\Box)F,
\end{equation}
where the appropriate gauge-fixing function is defined as
\begin{equation}
	\label{gff_2}
	F:=\bar{D}^2\left(v+4g^2\frac{1}{\Box f(\Box)}\bm{\bar{\phi}}\sigma_3\bm{\Phi}\right).
\end{equation}
This choice fixes the quantum gauge symmetry while preserving background invariance. Moreover, it cancels the unwanted mixed terms, as can be seen by substituting (\ref{gff_2}) into (\ref{gft_2}) and expanding the product:
\begin{equation}
	\label{gf_2}
	S_{GF2}=-\int d^7z\left[\frac{1}{4g^2}(D^2v)f(\Box)\bar{D}^2v+v\bm{\bar{\phi}}\sigma_3\bm{\Phi}+v\bm{\bar{\Phi}}\sigma_3\bm{\phi}+\bm{\bar{\phi}}\bm{M}^2\frac{1}{\Box f(\Box)}\bm{\phi}\right],
\end{equation}
where we have introduced the background-superfield-dependent mass matrix
\begin{equation}
	\bm{M^2}:=4g^2\sigma_3\bm{\Phi}\bm{\bar{\Phi}}\sigma_3=4g^2\left(\begin{array}{cc}
		\left|\Phi_+\right|^2 & -\Phi_+\bar{\Phi}_-\\
		-\Phi_-\bar{\Phi}_+ & \left|\Phi_-\right|^2
	\end{array}\right). 
\end{equation}
Now, since $M^2_v$ depends on the background matter superfields and $F$ changes under infinitesimal quantum gauge transformations as
\begin{equation}
	\delta_\Lambda F=i\bar{D}^2\left[\left(1-\frac{M^2_v}{\Box f(\Box)}\right)\bar{\Lambda}-4g^2\frac{1}{\Box f(\Box)}\left(\bar{\Lambda}\bm{\bar{\phi}}\right)\bm{\Phi}\right],
\end{equation}
it is clear that the ghosts interact with the background superfields through $M^2_v$ at the one-loop level. Thus, it is necessary to add to (\ref{quadratic_2}) the ghost action
\begin{align}
	S_{FP2}&=\left.\left(i\int d^5zc^\prime\delta_\Lambda F+i\int d^5\bar{z}\bar{c}^\prime\delta_\Lambda\bar{F}\right)\right|_{\Lambda\to c, \ \bar{\Lambda}\to \bar{c}}\\
	\label{ghost_2}
	&=\int d^7z\left[-c^\prime\left(1-\frac{M^2_v}{\Box f(\Box)}\right)\bar{c}+\bar{c}^\prime\left(1-\frac{M^2_v}{\Box f(\Box)}\right)c\right],
\end{align}
where we have omitted the cubic contributions, since they are irrelevant for one-loop calculations.

Finally, we can now collect the quadratic terms in the quantum superfields arising from (\ref{quadratic_2}), (\ref{gf_2}), and (\ref{ghost_2}) into a single functional expression:
\begin{equation}
	\begin{split}
		\label{combined_2}
		S^{(2)}_2+S_{GF2}+S_{FP2}^{(2)}=&\frac{1}{2}\int d^7z\bigg[\frac{1}{8g^2}v\left(-\Box f(\Box)+M_v^2+mf(\Box)\bar D^\alpha D_\alpha\right)v\\
		&+\bm{\bar{\phi}}\left(\bm{1}-\frac{\bm{M^2}}{\Box f(\Box)}\right)\bm{\phi}-c^\prime\left(1-\frac{M^2_v}{\Box f(\Box)}\right)\bar{c}\\
		&+\bar{c}^\prime\left(1-\frac{M^2_v}{\Box f(\Box)}\right)c\bigg].
	\end{split}
\end{equation}
From this result, we infer that the Hessians corresponding to each sector are given by
\begin{align}
	\label{H_v}
	\mathcal{H}_v&=\frac{1}{4g^2}\left(-\Box f(\Box)+M_v^2+mf(\Box)\bar D^\alpha D_\alpha\right);\\
	\label{H_phi_2}
	\mathcal{H}_\phi&=\left(\begin{array}{cc}
		\bm{0} & \displaystyle\left(\bm{1}-\frac{\bm{M^{2T}}}{\Box f(\Box)}\right)\bar{D}^2\\
		\displaystyle\left(\bm{1}-\frac{\bm{M^2}}{\Box f(\Box)}\right)D^2 & \bm{0}
	\end{array}\right);\\
	\label{H_FP2}
	\mathcal{H}_{FP2}&=\left(1-\frac{M^2_v}{\Box f(\Box)}\right)\left(\begin{array}{cc}
		\bm{0} & -\sigma_1\bar{D}^2\\
		\sigma_1 D^2 & \bm{0}
	\end{array}\right).
\end{align}
In the same manner as in the $\mathcal{N}=1$ case, the Hessian corresponding to the matter sector (\ref{H_phi_2}) is modified by the HD operator $f(\Box)$, even though we have not introduced it into the matter sector of (\ref{model_2}).

With the Hessians in hand, the next step is to obtain the one-loop contribution to the Euclidean effective action, which can be calculated from
\begin{equation}
	\label{1loopEA_2}
	\Gamma^{(1)}_{\mathcal{N}=2}=-\frac{1}{2}\textrm{Tr}\ln\mathcal{H}_v-\frac{1}{2}\textrm{Tr}\ln\mathcal{H}_\phi+\frac{1}{2}\textrm{Tr}\ln\mathcal{H}_{FP2}.
\end{equation}
The last two functional traces on the right-hand side of (\ref{1loopEA_2}) are identical to those found in \cite{Meu}. Thus, we can use Eq. (33) of \cite{Meu} to obtain
\begin{equation}
	\label{two_traces}
	-\frac{1}{2}\textrm{Tr}\ln\mathcal{H}_\phi+\frac{1}{2}\textrm{Tr}\ln\mathcal{H}_{FP2}=\frac{1}{2}\text{Tr}_+\ln\left(\Box f(\Box)-M^2_v\right)+\frac{1}{2}\text{Tr}_-\ln\left(\Box f(\Box)-M^2_v\right),
\end{equation}
where $\text{Tr}_+$ and $\text{Tr}_-$ denote the trace operations restricted to the chiral and antichiral subspaces, respectively.

The trace over the chiral subspace is evaluated by extracting the background-independent factor $\text{Tr}_+\ln\left(\Box f(\Box)\right)$ from the trace and subsequently discarding it, so that
\begin{equation}
	\label{chiraltrace}
	\begin{split}
		\text{Tr}_+\ln\left(\Box f(\Box)-M^2_v\right)&=\text{Tr}_+\ln\left(1-\frac{M^2_v}{\Box f(\Box)}\right)\\
		&=-\int d^7z\int \frac{d^3p}{(2\pi)^3}\frac{1}{p^2}\ln\left(1+\frac{M^2_v}{p^2 f(-p^2)}\right),
	\end{split}
\end{equation}
where the details of the calculation have been omitted, since it proceeds in a way very similar to that presented in \cite{BMS}. Additionally, the trace over the antichiral subspace is evaluated in an analogous way, yielding the same result as in (\ref{chiraltrace}). Therefore, by substituting (\ref{chiraltrace}) into (\ref{two_traces}), we obtain
\begin{equation}
	\label{two_traces_result}
	-\frac{1}{2}\textrm{Tr}\ln\mathcal{H}_\phi+\frac{1}{2}\textrm{Tr}\ln\mathcal{H}_{FP2}=-\int d^7z\int \frac{d^3p}{(2\pi)^3}\frac{1}{p^2}\ln\left(1+\frac{M^2_v}{p^2 f(-p^2)}\right). 
\end{equation}
Now we are left with the calculation of the first trace in (\ref{1loopEA_2}), which is explicitly given by
\begin{equation}
	-\frac{1}{2}\textrm{Tr}\ln\mathcal{H}_v=-\frac{1}{2}\textrm{Tr}\ln\left[\frac{1}{4g^2}\left(-\Box f(\Box)+M^2_v+mf(\Box)\bar D^\alpha D_\alpha\right)\right].
\end{equation}
Note that, if we disregard the Chern-Simons term by taking $m=0$, this contribution vanishes due to the absence of spinor covariant derivatives (this also occurs in the $\mathcal{N}=1$ case [see Eq. (\ref{1loopEA_1_2})]). Indeed, the absence of spinor covariant derivatives in $\textrm{Tr}\ln\left[\frac{1}{4g^2}\left(-\Box f(\Box)+M^2_v\right)\right]$ allows us to extract this term from the trace and subsequently discard it, so that we are left with
\begin{equation}
	\label{trace_v}
	-\frac{1}{2}\textrm{Tr}\ln\mathcal{H}_v=-\frac{1}{2}\textrm{Tr}\ln\left(1+\frac{mf(\Box)}{-\Box f(\Box)+M^2_v}\bar D^\alpha D_\alpha\right).
\end{equation}
In order to evaluate this trace, we need to use the following identity:
\begin{equation}
	\label{identity}
	\textrm{Tr}\ln\left(1+A(\Box)\bar D^\alpha D_\alpha\right)=\int d^7z\int \frac{d^3p}{(2\pi)^3}\frac{1}{p^2}\ln\left(1+p^2A^2(-p^2)\right).
\end{equation}
To prove it, we first use the definition of the trace of differential operators acting on superfields together with the series expansion of the logarithm. Thus,
\begin{equation}
	\label{explicity}
	\textrm{Tr}\ln\left(1+A(\Box)\bar D^\alpha D_\alpha\right)=\int d^7zd^7z^\prime\delta^7(z^\prime,z)\sum_{n=1}^{\infty}(-1)^{n+1}\frac{A^n(\Box)}{n}\left(\bar D^\alpha D_\alpha\right)^n\delta^7(z,z^\prime).
\end{equation}
Since
\begin{align}
	\left(\bar D^\alpha D_\alpha\right)^n&=-\Box^{\frac{n}{2}-1}D^\alpha \bar D^2 D_\alpha, \ \ \text{if $n$ is even};\\
	\left(\bar D^\alpha D_\alpha\right)^n&=\Box^{\frac{n-1}{2}}\bar D^\alpha D_\alpha, \ \ \text{if $n$ is odd},
\end{align}
and
\begin{equation}
	\delta^4(\theta^\prime,\theta)\bar D^\alpha D_\alpha\delta^7(z,z^\prime)=0 \ \ ; \ \ \delta^4(\theta^\prime,\theta)D^\alpha\bar D^2 D_\alpha\delta^7(z,z^\prime)=2\delta^7(z,z^\prime),
\end{equation}
only terms with even powers survive, and the expansion (\ref{explicity}) can be rewritten as
\begin{align}
	\textrm{Tr}\ln\left(1+A(\Box)\bar D^\alpha D_\alpha\right)&=\int d^7zd^7z^\prime\delta^3(x^\prime,x)\sum_{\ell=1}^{\infty}\frac{A^{2\ell}(\Box)}{\ell}\Box^{\ell-1}\delta^7(z,z^\prime)\\
	&=-\int d^7zd^3x^\prime\delta^3(x^\prime,x)\frac{1}{\Box}\ln\left(1-\Box A^2(\Box)\right)\delta^3(x,x^\prime),
\end{align}
which, in momentum space, corresponds to (\ref{identity}), thereby completing the proof.

Using the identity (\ref{identity}) in (\ref{trace_v}), we obtain, after a straightforward algebraic manipulation,
\begin{equation}
	\label{v_trace_result}
	\begin{split}
		-\frac{1}{2}\textrm{Tr}\ln\mathcal{H}_v&=-\frac{1}{2}\int d^7z\int \frac{d^3p}{(2\pi)^3}\frac{1}{p^2}\ln\left[\left(p^2f(-p^2)+M^2_v\right)^2+m^2p^2f^2(-p^2)\right]\\
		&+\int d^7z\int \frac{d^3p}{(2\pi)^3}\frac{1}{p^2}\ln\left(p^2f(-p^2)+M^2_v\right).
	\end{split}
\end{equation}
Notice that, on the right-hand side, the second integral is essentially the integral in (\ref{two_traces_result}), but with the opposite sign. Thus, by substituting the contributions (\ref{two_traces_result}) and (\ref{v_trace_result}) into (\ref{1loopEA_2}) and canceling the integrals, we arrive at
\begin{equation}
	\Gamma^{(1)}_{\mathcal{N}=2}=-\frac{1}{2}\int d^7z\int \frac{d^3p}{(2\pi)^3}\frac{1}{p^2}\ln\left[\left(p^2f(-p^2)+M^2_v\right)^2+m^2p^2f^2(-p^2)\right].
\end{equation}
Since this result was obtained under the assumption (\ref{constr_2}), it follows immediately that the one-loop K\"{a}hler EP is given by
\begin{equation}
	\label{pot_integral_2}
	K^{(1)}_{\mathcal{N}=2}=-\frac{1}{2}\int \frac{d^3p}{(2\pi)^3}\frac{1}{s}\ln A(s),
\end{equation}
where $s:=p^2$ and
\begin{equation}
	\label{arg}
	A(p^2):=\left(p^2f(-p^2)+M^2_v\right)^2+m^2p^2f^2(-p^2).
\end{equation}
Fortunately, the Euclidean Feynman integral (\ref{pot_integral_2}) is easier to evaluate than (\ref{pot_integral}) because of the product rule for logarithms. Indeed, let us first suppose that $A(s)$ has roots $s_1,s_2,\ldots,s_n$, not necessarily distinct. Since $A(s)$ is a nonconstant polynomial, it follows from the fundamental theorem of algebra that $A(s)$ admits a unique factorization (up to the ordering of the factors) of the form \cite{Axler}:
\begin{equation}
	A(s)=c\prod_{j=1}^{n}(s-s_j),
\end{equation}
where $c$ is a constant. Now, by substituting the factorization above into (\ref{pot_integral_2}) and using the product rule for logarithms, we obtain a sum of Feynman integrals that can be evaluated exactly by means of dimensional regularization. Therefore, the one-loop K\"{a}hler effective potential for the $\mathcal{N}=2$ higher-derivative Maxwell-Chern-Simons theory (\ref{model_2}) is given by
\begin{equation}
	\label{final_2}
	K^{(1)}_{\mathcal{N}=2}=-\frac{1}{4\pi}\sum_{j=1}^n\sqrt{-s_j}.
\end{equation}
As in the $\mathcal{N}=1$ theory, this one-loop correction is highly sensitive to the number of degrees of freedom in the HD theory (\ref{model_2}), which, in turn, depends on the degree of the polynomial $f(\Box)$. Indeed, the number of roots of the polynomial $A(s)$ is directly proportional to the degree of the polynomial $f(\Box)$, that is, as the degree of $f(\Box)$ increases, the number of contributions of the form $\sqrt{-s_j}$ to the one-loop correction (\ref{final_2}) also increases. 

Let us conclude this section by considering a simple application of our general result. Suppose a theory without higher derivatives, that is, $f(\Box)=1$. It follows from (\ref{arg}) that
\begin{equation}
	\label{arg_2}
	A(s)=\left(s+4g^2\bm{\bar{\Phi}}\bm{\Phi}\right)^2+m^2s=(s-s_+)(s-s_-),
\end{equation}
where we have used (\ref{mass_gauge_2}). The roots of $A(s)$ are negative real numbers and are given by
\begin{equation}
	\label{roots_2}
	s_\pm=-\frac{1}{2}\left[m^2+8g^2\bm{\bar{\Phi}}\bm{\Phi}\pm m^2\sqrt{1+\frac{16g^2\bm{\bar{\Phi}}\bm{\Phi}}{m^2}}\right].
\end{equation}
Finally, by substituting (\ref{roots_2}) into (\ref{final_2}), we arrive at the following one-loop correction:
\begin{equation}
	\begin{split}
		\left.K_{\mathcal{N}=2}^{(1)}\right|_{f(\Box)=1}&=-\frac{1}{4\pi}\sqrt{\frac{1}{2}\left[m^2+8g^2\bm{\bar{\Phi}}\bm{\Phi}+ m^2\sqrt{1+\frac{16g^2\bm{\bar{\Phi}}\bm{\Phi}}{m^2}}\right]}\\
		&-\frac{1}{4\pi}\sqrt{\frac{1}{2}\left[m^2+8g^2\bm{\bar{\Phi}}\bm{\Phi}- m^2\sqrt{1+\frac{16g^2\bm{\bar{\Phi}}\bm{\Phi}}{m^2}}\right]}.
	\end{split}
\end{equation}
Similarly to the result found in the previous section, we could also choose $f(\Box)=1+\frac{\Box}{\Lambda^2}$ as an application of our general result, but this choice of HD operator would lead to a polynomial $A(s)$ of degree four, and its roots would be highly cumbersome to write explicitly. For this reason, we did not include such an example in the present discussion.

\section{Conclusions}\label{secIV}

We have defined two HD extensions of the MCS theory in $\mathcal{N}=1$ and $\mathcal{N}=2$ superspaces. This was achieved by introducing, only in the gauge sector, an HD operator chosen to be a polynomial function of the d'Alembertian of arbitrary degree. For these theories, we calculated the one-loop corrections to the superfield EP. In contrast to our previous studies \cite{GNP,GGNPS}, all quantum calculations were performed in an HD $R_\xi$ gauge, and the EPs were obtained by evaluating the functional traces directly. The gauge choice is rather relevant because the EP generally depends on the gauge-fixing procedure, although the values of the EP at its extrema are gauge independent \cite{gauge-dependence}.

The two main results of this paper are the one-loop corrections given in (\ref{final_1}) and (\ref{final_2}) for the superfield EPs corresponding to the $\mathcal{N}=1$ and $\mathcal{N}=2$ versions of the HD MCS theories defined in (\ref{model_1}) and (\ref{model_2}), respectively. The importance of these results is twofold. First, from a technical point of view, they provide closed analytical expressions for the EPs in terms of the roots of polynomial functions, which, in previous treatments \cite{GNP,GGNPS}, were either not obtained for the MCS case or were left in an integral representation. Second, from a physical point of view, the results show how the one-loop corrections are directly controlled by the degrees of freedom associated with the combined Maxwell, Chern-Simons, and higher-derivative terms, where the latter include, in general, Ostrogradsky ghosts. Since the EP is a useful tool for studying the vacuum properties of a quantum field theory, our results indicate that the extra degrees of freedom generated by the HD operator, including possible Ostrogradsky-type ghosts, may affect the quantum-corrected ground-state configuration.

The most natural next step in this work would be to calculate the two-loop corrections to the superfield EPs of the HD MCS theories studied here. However, when compared to supersymmetric quantum electrodynamics with higher derivatives investigated in \cite{Meu}, the calculations in the present case would be more technically intricate due to the more complicated structure of the propagators for the gauge superfields, which would be obtained by inverting the Hessians (\ref{H_a}) or (\ref{H_v}), depending on the theory under consideration. These propagators would lead to a larger number of two-loop supergraphs that would need to be computed than the number of supergraphs considered in \cite{Meu}. For this reason, we plan to address this problem in future work, and the present contribution should be regarded as a first step in this direction.
 
\vspace{5mm}

{\bf Acknowledgments.} The author would like to thank A. Freitas for valuable discussions.

\end{document}